\documentclass{emulateapj}
\usepackage{apjfonts}

\usepackage{graphicx,amssymb,amsmath,times}

\newcommand{\lsi}    {\object{LS~I~+61~303}}
\newcommand{\lsif}   {LS~I~+61~303}
\newcommand{\xmm}    {{\it XMM-Newton}}
\newcommand{\swift}  {{\it Swift}}

\slugcomment{To appear in The Astrophysical Journal Letters.}

\shorttitle{CORRELATED X-RAY AND VHE EMISSION IN \lsi}
\shortauthors{H. Anderhub et~al.}

\begin{document}

\title{Correlated X-ray and Very High Energy emission in the gamma-ray binary \lsi}

%
\author{
H.~Anderhub$^{1}$,
L.~A.~Antonelli$^{2}$,
P.~Antoranz$^{3}$,
M.~Backes$^{4}$,
C.~Baixeras$^{5}$,
S.~Balestra$^{3}$,
J.~A.~Barrio$^{3}$,
D.~Bastieri$^{6}$,
J.~Becerra Gonz\'alez$^{7}$,
J.~K.~Becker$^{4}$
% ...
}
\address{$^{1}$ ETH Zurich, CH-8093 Switzerland}
\address{$^{2}$ INAF National Institute for Astrophysics, I-00136 Rome, Italy}
\address{$^{3}$ Universidad Complutense, E-28040 Madrid, Spain}
\address{$^{4}$ Technische Universit\"at Dortmund, D-44221 Dortmund, Germany}
\address{$^{5}$ Universitat Aut\`onoma de Barcelona, E-08193 Bellaterra, Spain}
\address{$^{6}$ Universit\`a di Padova and INFN, I-35131 Padova, Italy}
\address{$^{7}$ Inst. de Astrof\'{\i}sica de Canarias, E-38200 La Laguna, Tenerife, Spain}
% ...

\begin{abstract}

The discovery of very high energy (VHE) gamma-ray emitting X-ray binaries has
triggered an intense effort to better understand the particle acceleration,
absorption, and emission mechanisms in compact binary systems, which provide
variable conditions along eccentric orbits. Despite this, the nature of some of
these systems, and of the accelerated particles producing the VHE emission, is
unclear. To answer some of these open questions, we conducted a multiwavelength
campaign of the VHE gamma-ray emitting X-ray binary \lsi\ including the MAGIC
telescope, \xmm, and \swift\ during 60\% of an orbit in 2007 September. We
detect a simultaneous outburst at X-ray and VHE bands, with the peak at phase
0.62 and a similar shape at both wavelengths. A linear fit to the simultaneous
X-ray/VHE pairs obtained during the outburst yields a correlation coefficient
of $r=0.97$, while a linear fit to all simultaneous pairs provides $r=0.81$.
Since a variable absorption of the VHE emission towards the observer is not
expected for the data reported here, the correlation found indicates a
simultaneity in the emission processes. Assuming that they are dominated by a
single particle population, either hadronic or leptonic, the X-ray/VHE flux
ratio favors leptonic models. This fact, together with the detected photon
indices, suggests that in \lsi\ the X-rays are the result of synchrotron
radiation of the same electrons that produce VHE emission as a result of
inverse Compton scattering of stellar photons.

\end{abstract}


\section{Introduction}
\label{introduction}

\lsi\ is one of the few X-ray binaries that have been detected in very high
energy (VHE) gamma rays (see, e.g., \citealt{paredes08} for a recent review).
It is a high-mass X-ray binary system located at a distance of 2.0$\pm$0.2~kpc
\citep{frail91}. The system contains a rapidly rotating early type B0\,Ve star
with a persistent equatorial decretion disk and mass loss, and a compact object
with a mass between 1 and 4~$M_\odot$ orbiting it every $\sim$26.5~d in an
eccentric orbit (see \citealt{casares05}; \citealt{grundstrom07};
\citealt{aragona09}, and references therein). \lsi\ was classified as a
microquasar based on structures detected from five to several tens of
milliarcseconds with the European VLBI Network (EVN) and MERLIN
(\citealt{massi04}, and references therein), although analysis of later EVN and
MERLIN data sets does not reveal the presence of such structures
\citep{dhawan06,albert08_lsi2006_mw}. Very Long Baseline Array (VLBA) images
with a resolution of 1~mas (2~AU at the source distance) obtained during a full
orbital cycle show an elongated structure that rotates as a function of the
orbital phase \citep{dhawan06}. Later VLBA images show repeating structures at
the same orbital phases, reinforcing the idea that the milliarcsecond structure
depends on the orbital phase \citep{albert08_lsi2006_mw}. This may be
consistent with a model based on the interaction between the relativistic wind
of a young non-accreting pulsar and the wind/decretion disk of the stellar
companion \citep{dubus06a}, as occurs in PSR~B1259$-$63 (but confining the
pulsar wind may be problematic; see \citealt{bogovalov08}).

\lsi\ shows periodic non-thermal radio outbursts on average every $P_{\rm
orb}$=26.4960$\pm$0.0028~d, with the peak of the radio emission shifting
progressively from phase 0.45 to 0.95, using $T_0$=JD~2,443,366.775, with a
modulation period of 1667$\pm$8~d \citep{gregory02}. According to the most
precise orbital parameters, the periastron takes place at phase 0.275 and the
eccentricity of the orbit is $0.537\pm0.034$ \citep{aragona09}.

\lsi\ has been observed several times in X-rays (see \citealt{smith09} and
references therein). It generally displays quasi-periodic X-ray outbursts, with
the maximum occurring between orbital phases 0.4 and 0.8
\citep{goldoni95,taylor96,paredes97,harrison00,esposito07}, although the lack
of a sensitive long-term monitoring has prevented to search for a super-orbital
modulation. The source also shows short-term variability on timescales of hours
\citep{sidoli06}.

At VHE gamma rays, \lsi\ has been clearly detected several times both by MAGIC
\citep{albert06_lsi_science,albert08_lsi2006_mw,albert09_lsi2006_period} and
VERITAS \citep{acciari08_lsi,acciari09_lsi}. The source also displays VHE
gamma-ray periodicity, with minima taking place near periastron, where only
upper limits were found in MAGIC observations, and maxima occurring on average
at phase 0.6--0.7, although the source has also shown a second peak at phase
0.84 in one of the cycles \citep{albert09_lsi2006_period}. There are
indications of correlated X-ray/VHE emission, based on non-simultaneous data
taken more than 6~hr apart \citep{albert09_lsi2006_period}, and 1~d apart
\citep{albert08_lsi2006_mw}.

The lack of a systematic behavior from cycle to cycle at X-ray and VHE bands,
and the occurrence of short-term variability in the X-ray flux, has hampered
the definitive detection of an X-ray/VHE correlation from the comparison of
non-simultaneous data. We therefore conducted a multiwavelength campaign in
2007 September, covering the epoch of maximum VHE emission. Here, we report the
first simultaneous VHE and X-ray observations of \lsi\ obtained with the MAGIC
Cherenkov telescope and the \xmm\ and \swift\ X-ray satellites that show
correlated emission in both energy bands. We also briefly comment on radio and
optical spectroscopic observations obtained during the campaign.

\section{Observations and data analysis}
\label{observations}

\subsection{VHE Gamma Rays} \label{obs_vhe}

The VHE gamma-ray observations were performed from 2007 September 4 to 21 using
the MAGIC telescope on the Canary Island of La Palma ($28.75^\circ$N,
$17.86^\circ$W, 2225~m a.s.l.), from where \lsi\ is observable at zenith angles
above $32^\circ$. The essential parameters of MAGIC are a 17~m diameter
segmented mirror of parabolic shape (currently the largest dish for an air
Cherenkov telescope), an $f/D$ of 1.05, and an hexagonally shaped camera of 576
hemispherical photo multiplier tubes with a field of view of $3.5^\circ$
diameter. MAGIC can detect gamma rays from 60~GeV to several TeV, and with a
special setup the trigger threshold can be reduced to 25~GeV
\citep{aliu08_crab_science}. Its energy resolution is $\Delta E/E=20$\% above
energies of 200~GeV. The current sensitivity is 1.6\% of the Crab Nebula flux
for a $5\sigma$ detection in 50~hr of on-source time. The improvement compared
to the previous sensitivity was achieved by installing new 2~GHz FADCs
\citep{albert08_signal}.

The total observation time was 58.8~hr, including 39.6~hr in dark conditions
and 19.2~hr under moonlight or twilight \citep{britzger09}. The range of zenith
angles for these observations was $[32^{\circ},50^{\circ}]$, with 96\% of the
data having zenith angles below $43^{\circ}$. The final effective total
observation time is 54.2~hr. Table~\ref{table:vhe} gives the MJD, the effective
observation time, and the orbital phase for each of the MAGIC observations. The
observations were carried out in wobble mode \citep{fomin94}, i.e., by
alternately tracking two positions at $0.4^\circ$ offset from the source
position. This observing mode provides a reliable background estimate for
point-like sources such as \lsi.

\begin{table}
\begin{center}
\caption{Log of the VHE Gamma-ray Observations\label{table:vhe}}
\begin{tabular}{cccc}
\tableline\tableline
MJD$^a$  & Obs. Time & Phase Range  & $N (>300~{\rm GeV})$$^b$\\
         & (minutes) &              & (10$^{-12}$ cm$^{-2}$ s$^{-1}$)\\
\tableline
$54348.148\pm0.087$ & 203.0 & 0.470--0.476 & $\phantom{1}1.1 \pm 1.7$\\
$54349.157\pm0.080$ & 210.1 & 0.508--0.514 & $          -1.9 \pm 2.0$\\
$54350.136\pm0.094$ & 214.5 & 0.544--0.551 & $\phantom{1}0.2 \pm 2.1$\\
$54351.160\pm0.080$ & 219.8 & 0.584--0.590 & $          -0.1 \pm 2.2$\\
$54352.161\pm0.081$ & 221.1 & 0.621--0.627 & $          15.7 \pm 2.4$\\
$54353.163\pm0.082$ & 224.2 & 0.659--0.665 & $\phantom{1}7.7 \pm 2.1$\\
$54354.166\pm0.078$ & 212.5 & 0.697--0.703 & $\phantom{1}2.8 \pm 2.2$\\
$54355.153\pm0.091$ & 172.4 & 0.734--0.741 & $\phantom{1}0.3 \pm 2.3$\\
$54356.139\pm0.080$ & 148.7 & 0.771--0.777 & $\phantom{1}0.1 \pm 2.5$\\
$54357.153\pm0.092$ & 178.1 & 0.809--0.816 & $\phantom{1}6.4 \pm 2.3$\\
$54358.155\pm0.091$ & 179.3 & 0.847--0.854 & $\phantom{1}8.1 \pm 2.4$\\
$54359.154\pm0.092$ & 184.4 & 0.885--0.892 & $\phantom{1}2.9 \pm 2.4$\\
$54360.154\pm0.091$ & 177.3 & 0.923--0.929 & $\phantom{1}5.7 \pm 2.5$\\
$54361.154\pm0.094$ & 183.1 & 0.960--0.967 & $\phantom{1}5.3 \pm 2.4$\\
$54362.153\pm0.094$ & 188.7 & 0.998--0.005 & $          -0.6 \pm 2.4$\\
$54363.156\pm0.093$ & 138.6 & 0.036--0.043 & $\phantom{1}7.5 \pm 2.8$\\
$54364.149\pm0.099$ & 195.7 & 0.073--0.081 & $\phantom{1}3.8 \pm 1.9$\\
\tableline
\end{tabular}
\tablecomments{$^a$ The central MJD and half-span of each observation is quoted. $^b$ Uncertainties are given at 68\% confidence level.}
\end{center}
\end{table}

The data analysis was performed using the standard MAGIC analysis and
reconstruction software \citep{albert08_crab}. The quality of the data was
checked and data taken with anomalous event rates (very low atmospheric
transmission, car light flashes, etc.) were rejected following the standard
procedure, as previously done in \cite{albert09_lsi2006_period}. From the
remaining events, image parameters were calculated \citep{hillas85}. In
addition, the time parameters described in \cite{aliu09_timing} were calculated
as well. For the $\gamma$/hadron separation, a multidimensional classification
procedure based on the image and timing parameters with the Random Forest
method was used \citep{albert08_random}. We optimized the signal selection cuts
with contemporaneous Crab Nebula data taken at the same zenith angle range. The
energy of the primary $\gamma$-ray was reconstructed from the image parameters
using also a Random Forest method. The differential energy spectrum was
unfolded taking into account the full instrumental energy resolution
\citep{albert07_unfolding}. We estimate the systematic uncertainty to be about
30\% for the derived integral flux values and $\pm$0.2 for the obtained photon
index. For more details on the systematic uncertainties present in the MAGIC
data, see \cite{albert08_crab}.

\subsection{X-rays} \label{obs_xray}

We observed \lsi\ with \xmm\ during seven runs from 2007 September 4 to 11,
lasting from 12 to 18~ks (see Table~\ref{table:xray}), amounting to a total
observation time of 104.3~ks. The EPIC pn detector used the Large Window Mode,
while the EPIC MOS detectors used the Small Window Mode. All detectors used the
medium thickness optical blocking filter. Data were processed using version
8.0.0 of the \xmm\ Science Analysis Software (SAS). Known hot or flickering
pixels were removed using the standard SAS tasks. Further cleaning was
necessary to remove from the data set periods of high background corresponding
to soft proton flares, reducing the net good exposure durations to 67.0 and
92.6~ks for the pn and MOS detectors, respectively.

\begin{table*}[t!]
\begin{center}
\caption{Log of the \xmm\ (Top) and \swift\ (Bottom) X-ray Observations\label{table:xray}}
\begin{tabular}{cc@{~~~~}c@{~~~~}cc@{~~~~}c@{~~~~}ccr@{}l@{~~~~}r@{}l}
\tableline\tableline
          & \multicolumn{3}{c}{Exposure Time} \\
MJD$^a$   & Total & pn   & MOS  & Phase Range & $F (0.3-10~{\rm keV})$$^b$ & $\Gamma$ & $N_{\rm H}$ (\texttt{wabs})& $\chi^2$/ & d.o.f. & Variab./ & Signif.\\
          & (ks)  & (ks) & (ks) &             & ($10^{-12}$ erg~cm$^{-2}$~s$^{-1}$) & & ($10^{22}$~cm$^{-2}$) & & & (\%)/ & ($\sigma$)\\
\tableline
$54347.202\pm0.072$ & 12.4 &           10.8 &           12.4 & 0.434--0.440 & $13.8\pm0.2~~(\pm2.4)$ & $1.87\pm 0.03$ & $ 0.517\pm 0.014$ & 446.2/ & 382 &           17.1/ &           22.5\\
$54349.140\pm0.076$ & 13.1 & \phantom{1}5.6 & \phantom{1}9.1 & 0.507--0.513 & $12.4\pm0.2~~(\pm2.1)$ & $1.66\pm 0.03$ & $ 0.504\pm 0.019$ & 379.8/ & 352 &           16.8/ &           15.1\\
$54350.199\pm0.106$ & 18.3 &           10.5 &           17.1 & 0.546--0.554 & $13.3\pm0.2~~(\pm1.6)$ & $1.66\pm 0.02$ & $ 0.514\pm 0.013$ & 536.5/ & 422 &           11.8/ &           13.8\\
$54351.160\pm0.099$ & 17.1 &           10.9 &           12.1 & 0.583--0.590 & $13.4\pm0.2~~(\pm1.6)$ & $1.68\pm 0.03$ & $ 0.525\pm 0.015$ & 431.6/ & 400 &           12.1/ &           12.8\\
$54352.146\pm0.085$ & 14.7 & \phantom{1}5.8 &           13.2 & 0.621--0.627 & $22.9\pm0.2~~(\pm1.6)$ & $1.54\pm 0.02$ & $ 0.538\pm 0.012$ & 496.7/ & 447 & \phantom{1}6.8/ & \phantom{1}5.9\\
$54353.167\pm0.084$ & 14.5 &           11.9 &           14.5 & 0.659--0.665 & $18.6\pm0.2~~(\pm1.1)$ & $1.58\pm 0.02$ & $ 0.529\pm 0.013$ & 467.5/ & 429 & \phantom{1}5.9/ & \phantom{1}1.9\\
$54354.144\pm0.082$ & 14.2 &           11.5 &           14.2 & 0.696--0.702 & $12.6\pm0.2~~(\pm1.3)$ & $1.65\pm 0.03$ & $ 0.520\pm 0.015$ & 439.5/ & 407 &           10.0/ & \phantom{1}8.0\\
\tableline
$54354.663\pm0.073$ & \phantom{1}3.3 & ...  & ...            & 0.716--0.722 & $10.3\pm0.7~~(\pm2.0)$ & $1.71\pm0.16$  & 0.5 \phantom{11}(fixed)\phantom{1} &  5.38/& 14 &           18.4/ & \phantom{1}1.1\\
$54355.670\pm0.070$ & \phantom{1}2.7 & ...  & ...            & 0.754--0.759 & $12.4\pm0.9~~(\pm1.6)$ & $1.38\pm0.14$  & 0.5 \phantom{11}(fixed)\phantom{1} &  8.83/& 15 &           10.7/ & \phantom{1}0.3\\
$54356.671\pm0.073$ & \phantom{1}3.3 & ...  & ...            & 0.792--0.797 & $17.6\pm1.0~~(\pm2.0)$ & $1.36\pm0.10$  & 0.5 \phantom{11}(fixed)\phantom{1} &  9.83/& 26 & \phantom{1}9.9/ & \phantom{1}0.4\\
$54357.674\pm0.073$ & \phantom{1}3.3 & ...  & ...            & 0.830--0.835 & $13.9\pm0.9~~(\pm2.3)$ & $1.49\pm0.14$  & 0.5 \phantom{11}(fixed)\phantom{1} & 13.07/& 20 &           15.3/ & \phantom{1}1.2\\
$54358.178\pm0.103$ & \phantom{1}3.8 & ...  & ...            & 0.848--0.855 & $17.7\pm0.9~~(\pm1.5)$ & $1.47\pm0.09$  & 0.5 \phantom{11}(fixed)\phantom{1} & 20.46/& 32 & \phantom{1}6.3/ & \phantom{1}0.1\\
$54359.951\pm0.205$ & \phantom{1}5.3 & ...  & ...            & 0.911--0.926 & $19.5\pm0.8~~(\pm2.6)$ & $1.54\pm0.07$  & 0.5 \phantom{11}(fixed)\phantom{1} & 34.52/& 50 &           12.7/ & \phantom{1}0.4\\
$54362.247\pm0.237$ & \phantom{1}2.1 & ...  & ...            & 0.996--0.014 & $13.7\pm1.1~~(\pm3.6)$ & $1.78\pm0.19$  & 0.5 \phantom{11}(fixed)\phantom{1} &  5.45/& 11 &           25.0/ & \phantom{1}1.7\\
$54363.121\pm0.102$ & \phantom{1}2.5 & ...  & ...            & 0.034--0.042 & $12.3\pm1.0~~(\pm1.5)$ & $1.53\pm0.16$  & 0.5 \phantom{11}(fixed)\phantom{1} &  9.71/& 13 & \phantom{1}9.3/ & \phantom{1}0.1\\
$54365.530\pm0.040$ & \phantom{1}2.2 & ...  & ...            & 0.127--0.130 & $13.3\pm1.1~~(\pm1.7)$ & $1.46\pm0.15$  & 0.5 \phantom{11}(fixed)\phantom{1} & 10.14/& 12 &           10.1/ & \phantom{1}0.5\\
\tableline
\end{tabular}
\tablecomments{$^a$ The central MJD and half-span of each observation is quoted. $^b$ Fluxes are de-absorbed. The uncertainties in parentheses include the uncertainties due to intrinsic X-ray variability during the observations.}
\end{center}
\end{table*}

Cleaned pn and MOS event files were extracted for spectral analysis. Source
spectra were extracted from a $\sim$$70\arcsec$ radius circle centered on the
source (point-spread function of $15\arcsec$) while background spectra were
taken from a number of source-free circles with $\sim$$150\arcsec$ radius
(three for the pn detector, four for MOS1, and five for MOS2). The extracted
spectra were analyzed with XSpec v12.3.1 \citep{arnaud96}. The spectra were
binned so that a minimum of 20 counts per bin were present and energy
resolution was not oversampled by more than a factor 3. An absorbed power-law
function (\texttt{wabs*powerlaw}) yielded satisfactory fits for all
observations. De-absorbed fluxes in the 0.3--10~keV range were computed from
the spectral fits.

Additional observations of 2--5~ks each, consisting of several pointings of
0.2--1.0~ks, were obtained with the \swift/XRT from 2007 September 11 to 22.
The total observation time was 28.5~ks. The \swift\ data were processed using
the FTOOLS task {\tt xrtpipeline} (ver. 0.12.1 under HEASoft 6.6). The spectral
analysis procedures were the same as those used for the \xmm\ data. Since the
hydrogen column density was poorly constrained, with typical values of
$(0.5\pm0.2)\times10^{22}$~cm$^{-2}$, we fixed it to
$0.5\times10^{22}$~cm$^{-2}$, a typical value for \lsi\ also found in the \xmm\
fits and close to the average value for the \swift\ fits.

To search for short-term X-ray variability, we also extracted 0.3--10~keV
background-subtracted light curves for each observation, binned to 100~s for
\xmm\ and at half-time of the sparse 0.2--1.0~ks \swift\ pointings within each
observation. For \xmm, we considered the sum of the count rate in the three
detectors. From these light curves, we computed the degree of variability as
the standard deviation with respect to the mean of the count rate divided by
this mean, and the significance of this variability computed from the $\chi^2$
probability of the count rate being constant. We also computed hardness ratios
as the fraction between the count rates above and below 2~keV.

\section{Results}
\label{results}

\subsection{VHE Gamma Rays} \label{res_vhe}

\begin{figure}[t!]
\resizebox{1.0\hsize}{!}{\includegraphics[angle=0]{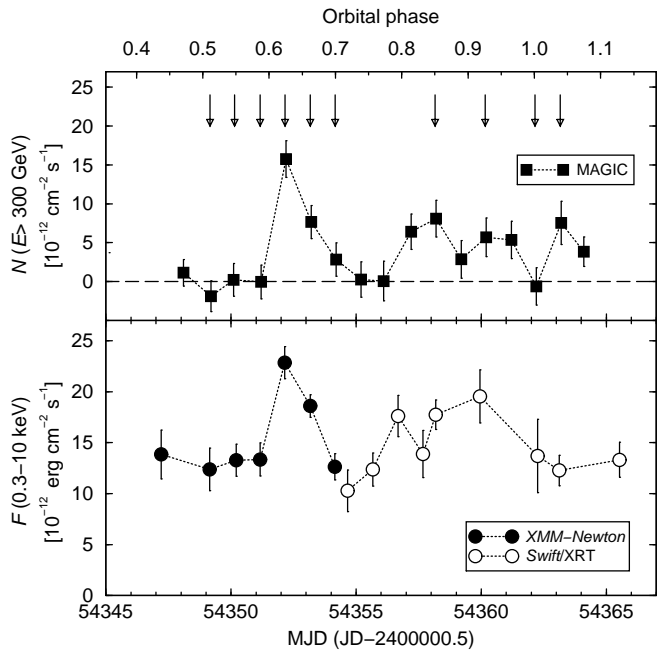}}
\caption{
VHE gamma-ray and X-ray light curves of \lsif\ during the multiwavelength
campaign of 2007 September. {\it Top}: flux above 300~GeV vs. the observation
time in MJD and the orbital phase. The horizontal dashed line indicates 0 flux.
The vertical arrows mark the times of simultaneous VHE gamma-ray and X-ray
observations. {\it Bottom}: de-absorbed flux in the 0.3--10~keV energy range
for the seven \xmm\ observations (filled circles) and the nine \swift\ ones
(open circles). Error bars correspond to a 1$\sigma$ confidence level in all
cases. Dotted lines join consecutive data points to help following the main
trends of the light curves. The sizes of the symbols are larger than the time
span of individual observations.
\label{fig:lc}}
\end{figure}

The measured fluxes of \lsi\ at energies above 300~GeV are listed in
Table~\ref{table:vhe}, and the corresponding light curve is shown in
Figure~\ref{fig:lc} (top). The periodically modulated peak emission
\citep{albert09_lsi2006_period} is prominently seen as the highest flux value
of these observations at phase 0.62 (followed by a smaller flux at phase 0.66).
In addition to this established feature we measured, as in 2006 December,
significant flux on a nightly basis at phase 0.85. We note that there is
significant flux in the phase range 0.8--1.0 at the level of
$(5.2\pm1.0)\times10^{-12}$~cm$^{-2}$~s$^{-1}$, which is compatible with the
2$\sigma$ upper limit we obtained for the shorter 2006 observations
\citep{albert09_lsi2006_period}.

We also determined the differential energy spectrum from the $\sim$10~hr of
observations conducted in the orbital phase range 0.6--0.7. A power-law fit yields:
\begin{align}
\frac{dN}{dE} = &\frac{(2.0 \pm 0.3_{\rm stat} \pm 0.6_{\rm syst})\times10^{-12}}{{\rm cm}^{2}~{\rm s}~{\rm TeV}} \nonumber \\ 
&\times \left(\frac{E}{1~{\rm TeV}}\right)^{-2.7 \pm 0.3_{\rm stat} \pm 0.2_{\rm syst}}, \nonumber
\end{align}
compatible within errors with our previous measurements
\citep{albert06_lsi_science, albert09_lsi2006_period}.

\subsection{X-rays} \label{res_xray}

We summarize in Table~\ref{table:xray} the parameters of the spectral fits and
variability obtained for both the \xmm\ and \swift/XRT data sets. All X-ray
fluxes quoted hereafter are de-absorbed. We note that there is no significant
hardness ratio change within each of the observations, in contrast to what was
found in the \xmm\ observations reported by \cite{sidoli06}. In principle this
implies that, for each observation, the flux and corresponding uncertainty
obtained from the spectral fit is a good estimate of the flux during the whole
observation. However, moderate count-rate variability on timescales of ks is
present in most observations, ranging from 6\% to 17\% for the \xmm\ data and
from 9\% to 25\% for the \swift\ data (see Table~\ref{table:xray}). The rms of
this variability should be considered in addition to the statistical
uncertainty when providing a flux measurement spanning several ks. We converted
this count-rate variability into flux variability by multiplying the degree of
variability defined in Section~\ref{obs_xray} with the fluxes coming from the
spectral fits. Since no spectral change is detected within each observation, we
added this flux variability in quadrature to the spectral fits flux errors.
This procedure provides the more realistic total flux uncertainties quoted in
parentheses in Table~\ref{table:xray}, and used hereafter.

We show in Figure~\ref{fig:lc} (bottom) the 0.3--10~keV light curve of \lsi.
The source displays a steady flux during the first four observations and shows
a steep increase at phase 0.62, which is followed by a slower decay up to phase
0.69 (\xmm) and probably up to phase 0.72 (\swift). The behavior is very
similar to the one seen at VHE gamma rays, although at X-ray energies the
baseline has a significant flux. Later on there is a significant increase of
the X-ray flux up to phase 0.8. This high flux is detected with a sparse
sampling up to phase 0.9, and the source goes back to its baseline flux at
phase 1.0. This high X-ray flux between phases 0.8 and 1.0 occurs when the
source is also detected at VHE gamma rays.

\subsection{X-Ray/VHE Gamma-Ray Correlation}

\begin{figure}[t!]
\resizebox{1.0\hsize}{!}{\includegraphics[angle=0]{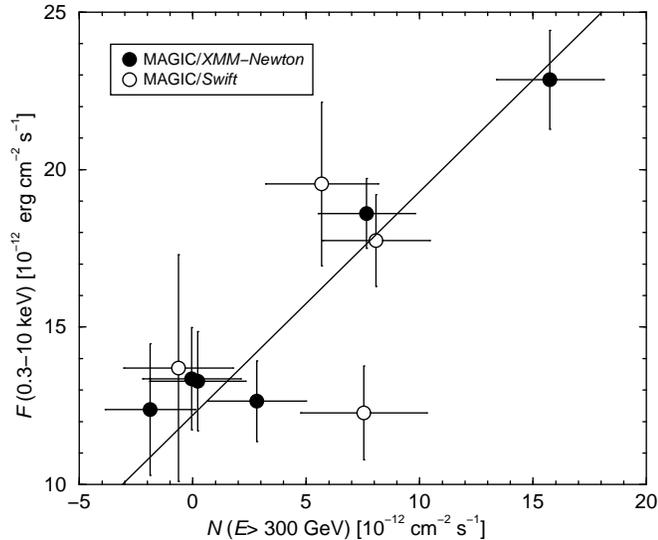}}
\caption{De-absorbed X-ray fluxes as a function of VHE gamma-ray fluxes. Only
the 10 simultaneous fluxes, marked with arrows in Figure~\ref{fig:lc}, have
been considered. Error bars correspond to a 1$\sigma$ confidence level in all
cases. The solid line represents a $\chi^2$ linear fit to all data points.
\label{fig:corr}}
\end{figure}

A clear correlation between the X-ray and VHE gamma-ray emissions is seen
during the outburst, with a simultaneous peak at phase 0.62 (see
Figure~\ref{fig:lc}). To study the significance of this correlation, we
selected the X-ray data sets that overlap with MAGIC observations. There are
six overlapping MAGIC/\xmm\ data sets, for which strictly simultaneous
observations range from 3.3 to 3.9~hr. For the four overlapping MAGIC/\swift\
data sets, the strictly simultaneous observations range from 2.2 to 4.1~hr
(although the \swift\ runs have gaps). We plot in Figure~\ref{fig:corr} the
X-ray fluxes against the VHE fluxes (from Tables~\ref{table:vhe} and
\ref{table:xray}) for all 10 simultaneous pairs, which are marked with arrows
in Figure~\ref{fig:lc}. The linear correlation coefficient for the six
simultaneous MAGIC/\xmm\ pairs that trace the outburst is $r=0.97$. For the ten
simultaneous pairs, we find $r=0.81$ (a $|r|$ larger than that has a
probability of about $5\times10^{-3}$ to be produced from independent X-ray and
VHE fluxes). Minimizing $\chi^2$ we obtain $\chi^2=7.68$ for 8 degrees of
freedom and the following relationship:
$F$(0.3--10~keV)/[10$^{-12}$~erg~cm$^{-2}$~s$^{-1}]=(12.2^{+0.9}_{-1.0})+(0.71^{+0.17}_{-0.14})\times
N$($E>$300~GeV)/[10$^{-12}$~cm$^{-2}$~s$^{-1}]$ (non-Gaussian uncertainties).
This fit is plotted as a solid line in Figure~\ref{fig:corr}.

However, as can be seen in Figure~\ref{fig:corr}, the flux uncertainties are
relatively large. This calls for a test of the reliability of the correlation
strength considering the errors of individual data points. To this end, we use
the $z$-Transformed Discrete Correlation Function (ZDCF), which determines 68\%
confidence level intervals for the correlation coefficient from unevenly
sampled data (see, e.g., \citealt{edelson88}; \citealt{alexander97}). The
Fisher $z$-transform of the linear correlation coefficient is used to estimate
the 68\% confidence level interval. Applying the ZDCF to the X-ray and VHE
light curves reported here we obtain the following uncertainties for the linear
correlation coefficient: $r=0.81_{-0.21}^{+0.06}$. The sensitivity of the MAGIC
telescope requires observation times of several hours to be able to get
few-sigma detections for the VHE fluxes from \lsi. During such long periods,
the X-ray fluxes have quite large intrinsic variability (see
Table~\ref{table:xray}). This results in quite large uncertainties on both the
VHE and X-ray fluxes that restrict the ability to determine a possible
correlation out of the outburst.

\section{Discussion}
\label{discussion}

We have discovered an X-ray/VHE gamma-ray correlation in the gamma-ray binary
\lsi\ based on simultaneous multiwavelength data obtained with MAGIC, \xmm, and
\swift\ during a single orbital cycle. The correlation is mainly due to the
very similar trends of the detected flux during the outburst around orbital
phase 0.62, while the uncertainties prevent to be sure about the existence of
the correlation outside the outburst. Given the variability in the X-ray flux
(up to 25\% on hour scales), it is necessary to program simultaneous
observations to perform correlation studies in \lsi. This conclusion has also
been reached recently by the VERITAS Collaboration, when reporting the lack of
X-ray/VHE correlation based on contemporaneous data with a VHE sampling that is
not dense enough \citep{acciari09_lsi}. Although a similar X-ray/VHE
correlation has been obtained for the VHE gamma-ray emitting X-ray binary
\object{LS~5039}, this result was based on non-simultaneous data acquired years
apart \citep{aharonian05_ls5039_science,aharonian06_ls5039_period,takahashi09}.

The VHE emission within a binary system can suffer photon--photon absorption
via pair creation, mainly with the stellar optical/ultraviolet photons. In
\lsi, this absorption is only expected to be significant toward the observer
for $E>300$~GeV just before periastron
\citep{dubus06b,bednarek06,sierpowska09}, a phase range not explored here. In
addition, the quoted X-ray fluxes are already de-absorbed (and the hydrogen
column densities and the associated errors are low). Overall, there are no
absorption effects to be considered. Therefore, the X-ray/VHE correlation we
have found for \lsi\ cannot be an artifact due to variable absorption toward
the source. This indicates that the emission processes at both wavelengths
occur at the same time and are probably the result of a single physical
mechanism. In this context, it is reasonable to assume that the X-ray and VHE
emissions are produced by a single particle population.

It is interesting to note that the MAGIC spectrum in the 0.6--0.7 phase range
yields $\sim11\times10^{-12}$~erg~cm$^{-2}$~s$^{-1}$ for $E>300$~GeV, while the
X-ray flux is $\sim19\times10^{-12}$~erg~cm$^{-2}$~s$^{-1}$. Therefore, the
total X-ray flux is approximately twice the VHE flux. However, if we subtract
an apparent baseline X-ray flux of $10\times10^{-12}$~erg~cm$^{-2}$~s$^{-1}$,
the resulting X-ray flux is similar to the total VHE flux in the phase range
0.6--0.7. If the radiation mechanisms are dominated by a single particle
population, the X-ray/VHE correlation and the smaller/similar VHE fluxes favor
leptonic models. In hadronic models, the X-ray emitting $e^\pm$ and the VHE
photons would come from the same protons (for reasonable values of the magnetic
field), and the luminosity of the $e^\pm$ radiation should be $\la 1/2$ that of
VHE gamma-rays (see Figure~5 of \citealt{kelner06} for reasonable proton energy
distributions) unlike it is observed. In addition, the inverse Compton (IC)
cooling channel is less efficient than the synchrotron channel to produce the
detected X-ray emission for reasonable values of the magnetic field (see
\citealt{takahashi09} for a similar discussion for \object{LS~5039}). This
clearly suggests that the X rays are the result of synchrotron radiation of the
same electrons that produce VHE emission as a result of IC scattering of
optical/ultraviolet stellar photons.

The observed photon indices of the simultaneous X-ray and VHE spectra are
consistent with one population of electrons following a power-law energy
distribution with index $\sim$2.1. These electrons would produce X-rays via
synchrotron and VHE photons via IC with an interaction angle $\la \pi/2$. We
note that an electron index of $\sim$2.1 is too hard if synchrotron cooling
dominates in the X-ray range, since it implies an injected electron index of
1.1. On the other hand, dominant adiabatic cooling implies an injection index
of 2.1, a more reasonable value (see, for example, the discussion in
\citealt{takahashi09}). Although small changes in the VHE spectrum would be
expected due to variations in the IC interaction angle or the electron index
(as seen in X-rays), at present they are not detectable due to the large
uncertainties of the VHE photon index.

Finally, we note that contemporaneous radio light curves obtained with RATAN,
VLBA images, and H$\alpha$ spectroscopy are consistent with previous results
\citep{gregory02,dhawan06,zamanov99}. Details on these observations will be
reported elsewhere. Therefore, the X-ray/VHE correlation occurred when the
source was showing a standard behavior in both its outflow (radio) and
decretion disk (H$\alpha$ line).

\acknowledgments

We thank the Instituto de Astrofisica de Canarias for the excellent working
conditions at the Observatorio del Roque de los Muchachos in La Palma. The
support of the German BMBF and MPG, the Italian INFN and Spanish MICINN is
gratefully acknowledged. This work was also supported by ETH Research Grant TH
34/043, by the Polish MNiSzW Grant N N203 390834, and by the YIP of the
Helmholtz Gemeinschaft.

{\it Facilities:} \facility{MAGIC}, \facility{{\it XMM}}, \facility{{\it
Swift}}, \facility{RATAN}, \facility{VLBA}, \facility{Skinakas: 1.3 m}

\end{document}